\documentclass[12pt, letterpaper]{article}

\usepackage{amsmath,amssymb}%Needed for multline
\usepackage{cite}%Needed for long citations
\usepackage{fancyhdr}%Needed for title page headings
\usepackage[top=1in, bottom=1in, left=1in, right=1in]{geometry}%Needed for margins
\usepackage{graphicx}%Needed for includegraphics
\usepackage{hyperref}%Needed for hyper references

\numberwithin{equation}{section}%Needed for equation numbers to follow sections
\date{}%Needed to remove the date

\title{{\rm\footnotesize \qquad \qquad \qquad \qquad \qquad \ \qquad \qquad \qquad \ \ \ \ \ \                  UTTG-08-18\ TCC-032-18     RUNHETC-2018-11     
}\vskip.5in    The Holographic Space-time Model of Cosmology\\ Winner 5th Prize 2018 Gravitation Research Foundation Essay Contest }
\author{Tom Banks (corresponding author)\\
NHETC and Department of Physics \\
Rutgers University, Piscataway, NJ 08854-8019\\
E-mail: \href{mailto:tibanks@ucsc.edu}{tibanks@ucsc.edu}
\\
\\
W. Fischler\\
Department of Physics and Texas Cosmology Center\\
University of Texas, Austin, TX 78712\\
E-mail: \href{mailto:fischler@physics.utexas.edu}{fischler@physics.utexas.edu}}
\begin{document}

\maketitle
\thispagestyle{fancy} %Needed to remove the page number from the title page

\begin{abstract}This essay outlines the Holographic Space-time (HST) theory of cosmology and its relation to conventional theories of inflation. The predictions of the theory are compatible with observations, and one must hope for data on primordial gravitational waves or non-Gaussian fluctuations to distinguish it from conventional models. The model predicts an early era of structure formation, prior to the Big Bang. Understanding the fate of those structures requires complicated simulations that have not yet been done. The result of those calculations might falsify the model, or might provide a very economical framework for explaining dark matter and the generation of the baryon asymmetry.

\normalsize \noindent  \end{abstract}

%%%%%%%%%%%%%%%%%%%%%%%%%%%%%%%%%%%%%%%%%%%%%%%%%%%%%%%%%%%%%%%%%%%%%%%%%%%%%%%%%%%%%%%%%%%%%%%%%%%%%%%%%%%%%%%%%%%%%%%%%%%%%%%%%%%%%%%%%%%%%%%%%%%%

\newpage
\tableofcontents
\vspace{1cm}

%%%%%%%%%%%%%%%%%%%%%%%%%%%%%%%%%%%%%%%%%%%%%%%%%%%%%%%%%%%%%%%%%%%%%%%%%%%%%%%%%%%%%%%%%%%%%%%%%%%%%%%%%%%%%%%%%%%%%%%%%%%%%%%%%%%%%%%%%%%%%%%%%%%%
\vfill\eject
\section{Holographic Space-time}

Leibniz' famous question, ``Why is there something rather than nothing?", can be viewed as the beginning of modern cosmology.  In the 19th century Boltzmann sharpened the puzzle of how the universe began, by noting that the universality of the second law of thermodynamics indicated that the universe started in a state of very low entropy.  His attempt to explain this as an accidental fluctuation, was quickly shot down by his assistant Schutz. In the 20th century, Penrose emphasized the importance of Boltzmann's question and tried to connect it to the vanishing of the Weyl tensor in early cosmological history.  The theory of Holographic Space-time (HST) sheds new light on these ancient questions.  It posits that ``nothing" is actually the state of maximal entropy of the universe, because in that state all degrees of freedom in the universe live on the cosmological horizon, with a dynamics that scrambles information at the maximal rate allowed by causality.  Localized excitations inside a causal diamond are constrained states of the holographic degrees of freedom.  The constraints guarantee that they decouple from most of the scrambled boundary variables for a long time.  The answer to both Boltzmann's and Leibniz' questions is that without a low entropy beginning, localized excitations (``something") will only exist as rare isolated thermal fluctuations.   Since localized excitations, whose interactions approximately obey the rules of local field theory, are crucial to our understanding of any kind of complex organization, this seems like a satisfactory answer to both questions.  It is a very weak kind of environmental selection criterion, which does not require us to have a detailed understanding of exobiology.

This essay will attempt to explain the (HST) model of cosmology\cite{holo}\cite{holoinflation2}\cite{holoinflation3} without technical details.  The model is an explicit collection of quantum mechanical systems but we will use only simple thermodynamic estimates to explain a picture of the evolution of the universe that can explain the gross features of cosmology in terms of a small number of parameters.  In addition we'll note that the same principles give a model of the interior of black holes, which does not suffer from the notorious ``firewall paradox".

A causal diamond in space-time is the region to which a detector on a time-like trajectory can send signals and receive a response to those signals, during an interval of proper time $T$.  A nested sequence of proper time intervals, which completely specifies the trajectory, is equivalent to a nested sequence of causal diamonds.  The boundary of a causal diamond is a null surface. The space-like area at a fixed null coordinate on the boundary grows as one moves away from both the tips of the diamond and takes on a maximum somewhere on the boundary.  The basic principle of HST is that the maximal area, in Planck units, divided by $4$ is equal to the logarithm of the dimension of the Hilbert space required to describe all the quantum information associated with the diamond.  The primary reason to believe this claim is Jacobson's\cite{ted} demonstration that the hydrodynamical equations of this entropy law, coupled with Unruh's formula for the temperature of an infinitely accelerated trajectory, are just Einstein's equations
\begin{equation} k^{\mu} k^{\nu} (R_{\mu\nu} - \frac{1}{2} g_{\mu\nu} R - 8 \pi G_N T_{\mu\nu}) = 0 , \end{equation}  where $k^{\mu}$ is an arbitrary null tangent vector field. The only term in Einstein's equations that is not derivable from hydrodynamics is the cosmological constant (c.c.).  Thus, this parameter should not be thought of as a local contribution to the energy density. Banks and Fischler\cite{tbwf} conjectured that the c.c. was an asymptotic boundary condition, controlling the growth of area for large proper time.  In the case of future asymptotically dS space, with positive c.c., the area approaches a finite constant in the limit of infinite proper time.

In HST, quantum dynamics is done independently along each time-like trajectory.  The proper time Hamiltonian is time dependent, and during each interval of time it couples together only degrees of freedom (DOF) associated with the corresponding causal diamond by the entropy law.  Note that even in Minkowski space the vector field, which moves time slices causally related to a system at the tips of each diamond, into each other, is not a Killing vector, so we expect a time dependent Hamiltonian. 

The intersection between causal diamonds on two different trajectories, contains a maximal causal diamond, which should be viewed as a subsystem\footnote{In quantum mechanics, a subsystem is a tensor factor in the full Hilbert space.} of both systems, associated with the individual trajectories.  This is implemented by insisting that the density matrices assigned to that subsystem by the dynamics of each trajectory, have the same spectrum.  The time slices associated with proper time dynamics interpolate between causal diamonds. In cosmology, they coincide with FRW time slices only at the point where the trajectory crosses the FRW slice.  At other points they go back to the FRW past.  
\begin{figure}[htb]
\begin{center}
  \includegraphics[width=0.80\textwidth]{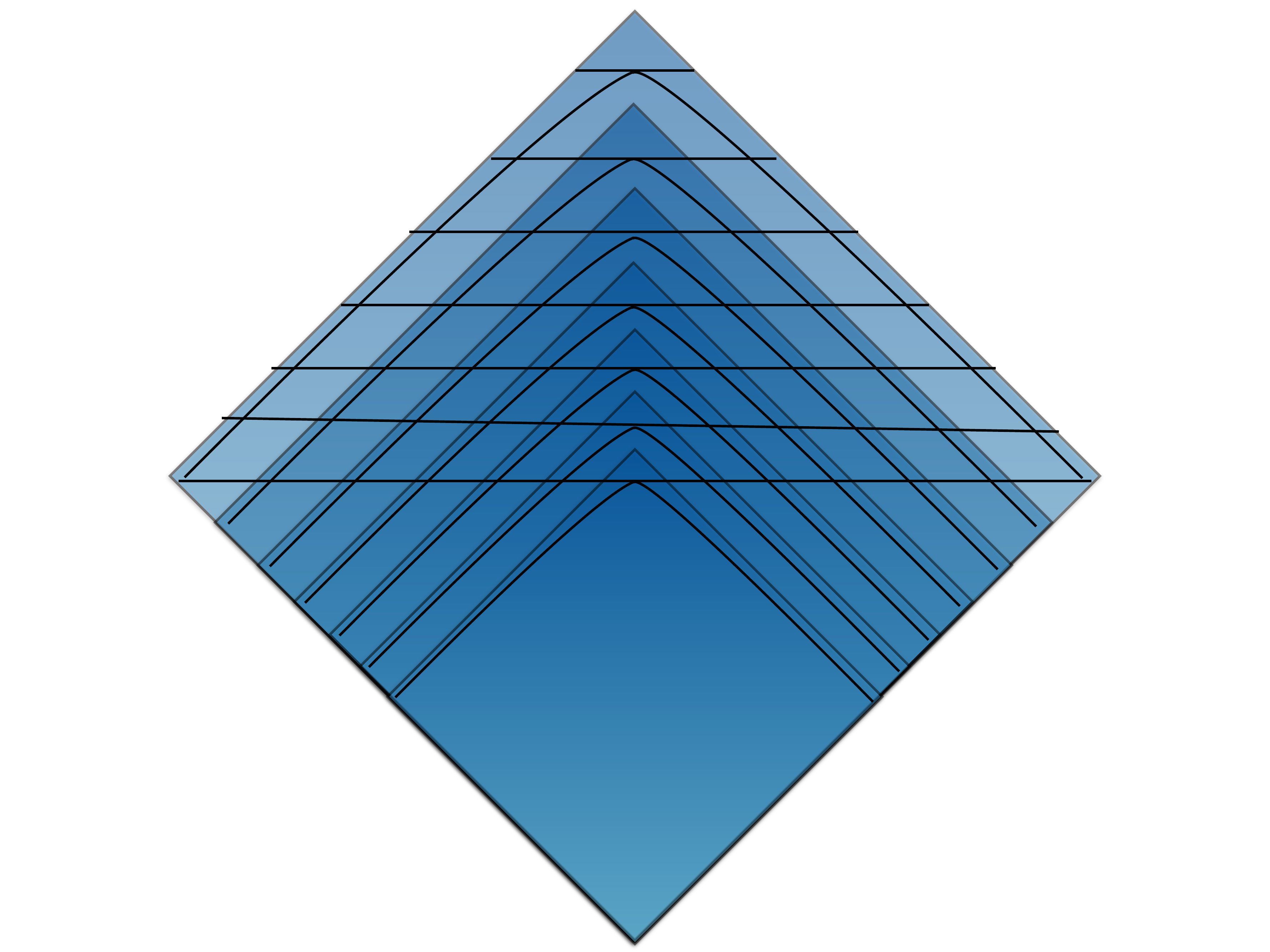}
   \caption{Horizontal slices are FRW, hyperbolic slices are HST.\label{fig1}}
   \end{center}
\end{figure}

One of the most important principles of HST can be understood by examining the metric of a localized object in dS space.   It is
\begin{equation} ds^2 = - (1 - R_S/r - r^2/R^2) dt^2 + \frac{dr^2}{1 - R_S/r - r^2/R^2} + r^2 d\Omega^2 . \end{equation}  $R$ is the Hubble radius of dS space and $R_S \equiv 2 G_N M $ the Schwarzschild radius of the object, which we assume $ \ll R$.  Using the equivalence between horizon area and entropy, one finds that the localized object is a constrained subsystem of the system described by the empty dS metric, with an entropy deficit corresponding to a Boltzmann factor with precisely the Gibbons-Hawking temperature of dS space. This derivation of the Gibbons-Hawking temperature does not require quantum field theory.

For a given entropy deficit $M/T$, the maximal entropy subsystem is a localized black hole.  The principles of HST assert that something similar is true for any causal diamond: objects localized in the bulk of the diamond correspond to low entropy constrained states of the fundamental DOF.  Applied to Minkowski space this leads to a theory of scattering of jets of particles, in which black hole like resonances form and dissipate in a manner completely consistent with unitarity\cite{holonewton}\footnote{The extant models of Minkowski space are incomplete, because they do not incorporate the constraint of Lorentz invariance, although they do implement unitarity and causality.}.  The temperature of black holes is explained in the same way that we explained the temperature of dS space.

\section{HST Cosmology}

Now consider the evolution of the universe, starting from some finite time $t = 0$ along a given timelike trajectory. This is the point where the trajectory touches the Big Bang hypersurface of the FRW model, which describes the hydrodynamics of this system. Assume at later times $t$ that the entropy(area) grows with $t$ until it reaches some maximum of order $m^2 \gg 1$. By this we mean not only that the size of the Hilbert space associated with the causal diamond from $0$ to $t$ is growing, but that the actual state of the system is random.  Initially, we have an area that grows like $t^2$.  A period of inflation would correspond to a relatively long interval of proper time during which the area remains constant.  That would correspond to a causal patch of dS space with Hubble radius $R_{dS} = m L_P$.  Up to this point in time, the system has no localized objects in it, because it always has maximal entropy.  The space time scale factor $a(t) = \sinh^{1/3} ({3t/R_{dS}})$, describes the coarse grained thermodynamics of this universe, whose perfect fluid stress tensor is a mixture of $ p = \rho$ and $p = - \rho$.  One can model the $p = \rho$ equation of state in terms of a collection of fields with no spatial derivative terms, but that model would not have finite entropy in causal diamonds.  It does serve to demonstrate that such a universe has no propagating excitations.  

We now assume that after the inflationary period\footnote{We'll understand below why there {\it must} be an inflationary  period.}, as the area of the apparent horizon expands, we encounter states of less than maximal entropy in the larger Hilbert space.   The lesson we learned from the dS black hole entropy formula tells us that such states will have an entropy deficit that scales like the square root of the area of the cosmological horizon, $c \sqrt{A}/L_P$.  For fixed coefficient $c$, the maximal entropy state will consist of a collection of localized black holes.  This will evolve into a collection of localized black holes in the asymptotic dS space, plus radiation coming from the Hawking decay of the black holes.  

If the collection of black holes is very inhomogeneous, it will evolve, for any given FRW geodesic, into a collection of large black holes bound to the geodesic, plus some that are swept out of the horizon.  At no point will the universe be filled with an approximately uniform gas of radiation.  The Hawking radiation from black holes bound to the geodesic will dribble out slowly and then suddenly explode.

We conjecture that the most probable way to get a conventional radiation dominated universe is to assume that the black holes that come into the horizon form a maximally uniform dilute gas, in which density fluctuations are initially small.  The gas cannot be exactly uniform because each black hole is a finite quantum mechanical system.  Properties like the black hole mass and angular momentum are averages, subject to fluctuations of order $ 1/\sqrt{entropy} \sim 1/m . $  $m$ must be large in order for the fluctuations to be small.  The usual assumptions of statistical mechanics tell us that non-Gaussian fluctuations are suppressed by powers of $1/m$. $m$ must be $\ll R$ ($R\sim 10^{61}$ is the radius of the cosmological horizon in Planck units) in order for the intermediate era of the universe with localized excitations to consist of something other than a few black holes that almost fill the observable universe.  While such large black holes eventually decay to radiation, the decay time is much longer than the time for the emitted radiation to exit the cosmological horizon.

This leads to predictions for two and three point fluctuations of the space-time metric in comoving coordinates : \begin{equation} \langle \zeta (k) \zeta (-k) \rangle = \frac{C_1}{m^2 \epsilon^2 (t_k)} S_2 (k),  \end{equation}  \begin{equation} \langle \gamma (k) \gamma (-k) \rangle = \frac{C_2}{m^2} T_2 (k),  \end{equation}
\begin{equation} \langle \zeta (k) \zeta (l) \zeta (- k - l) \rangle = \frac{C_2}{m^3 \epsilon^2 (t_k) } S_3 (k,l),  \end{equation}
\begin{equation} \langle \zeta (k) \gamma (l) \gamma (- k - l) \rangle = \frac{C_3}{m^3  } [TS]_3 (k,l) , \end{equation}
\begin{equation} \langle \gamma (k) \gamma (l) \gamma (- k - l) \rangle = \frac{C_4}{m^3 } T_{31} (k,l) + \frac{C_5}{m^3 } T_{32} (k,l) + \frac{C_6}{m^3 } T_{33} (k,l) . \end{equation}  Here $S_2 , S_3, [TS]_3,$ and $T_{3i}$, are the standard dS invariant form factors with conventional normalization and the $C_i$ are constants that are not yet calculable in the HST formalism.  We've used Maldacena's squeezed limit theorems and $SO(1,4)$ invariance to estimate the factors of $\epsilon$.

These are significantly different than the predictions from slow roll inflation, despite the fact that, as we will see, both formalisms invoke an inflationary era followed by a period of slow roll.   For the two point function of scalar fluctuations, the two theories predict a different form as a function of the slow roll parameter at horizon crossing $\epsilon (t_k) = \dot{H}/H^2 $. However, since our only present observational constraint on the slow roll metric comes from CMB measurements of the scalar two point function, both theories can explain the data.  

The two theories give different predictions for the tilt in the primordial tensor spectrum, but the differences are proportional to $\epsilon$ and small.  Perhaps more significant is the fact that the magnitude of the tensor to scalar ratio is $16\epsilon$ in field theoretic inflation and $\sim \epsilon^2$ in HST.  This suggests that this ratio should be smaller in HST models than field theory based inflation models.  Unfortunately the unknown coefficients $C_{i}$ in the HST formulae, combined with the factor of $16$ in the field theory calculation (which suggests that these calculations can contain large natural numbers $\sim 10^2$), and the fact that $\epsilon \sim 0.1$, do not allow one to make the definite conclusion that primordial tensor fluctuations should be unobservable in the HST model.

The situation is complicated by the fact that the HST model has a secondary source of gravitational wave fluctuations, coming from the decay of the black holes, the fluctuations in whose masses are the source of scalar perturbations. This source of tensor fluctuations has the same tilt as the scalar fluctuations but it is smaller by a factor of $1/g$ where $g$ is the number of effectively massless particle states at the reheat temperature.
The reheat temperature is simply determined by the energy density of black holes at the time the black holes decay.  That energy density is initially of order $m^{-2} $ and it dilutes to $m^{- 6}  = 10^{40}$ GeV$^4$ (recall that throughout this essay we've worked in Planck units).
Thus $T_{RH} \sim 10^9 - 10^{10} $ GeV.  $g$ is thus likely to be of order $10^3$, particularly in particle physics models incorporating supersymmetry below the reheat scale.  

The HST model predicts that the $k$- point functions of fluctuations are approximately $SO(1,4)$ invariant if the number of e-folds of inflation is large, and it predicts $N_e \sim 6	0$.  Without entering into too much technical detail, the basic reason for the invariance is that, within the HST formalism, dS space is the natural endpoint of evolution of a universe that spends most of its post Planck scale history in a phase with equation of state $p = \rho$.  Causal diamonds in that universe have no local excitations and the dynamics on their boundary at time $t$ is described by a cutoff $1 + 1$ dimensional conformal field theory, with central charge $c \sim t^2$.  The dS endpoint is simply the result of restricting the Hilbert space to a finite dimension, which is equivalent to stopping the growth of the central charge.  The model thus has an {\it approximate} $SL(2,R)$ symmetry.  In the HST inflation model, each black hole is described by such an approximate CFT, and the distribution of black holes is invariant under the $SO(3)$ group of rotations around the particular trajectory under discussion.  One then constructs the full (approximate) $SO(1,4)$ generators by integrating over the local $SL(2,R)$ generators at different points on the celestial sphere.  

As a consequence of this symmetry, the three point correlation functions have the forms written above.  The tensor three point function is interesting because the two parity conserving form factors are of roughly the same order of magnitude.  In field theoretic inflation $C_{5} \sim H/m_P C_{4} $ and the parity violating term $C_6$ vanishes to all orders in $H/m_P$ since it is a phase in the Wheeler-DeWitt wave function, and the commuting operators that give the correlation functions, are insensitive to it.  Since the universe does not appear to have any exact reflection symmetry, we expect that there is no such all orders suppression in the HST model.   Unfortunately, the prospects of measuring the tensor three point function are extremely remote.

To understand why the HST model is a model of inflation, one has to go back to the picture relating FRW time slices to the causal time slices used in HST.  In the previous discussion we've used the FRW description, but the quantum dynamics of the model uses causal time slices.  A black hole that enters into the apparent horizon of a particular trajectory at some causal time $t$, is sitting on an FRW slice at an earlier time, and a different position.  There is a timelike FRW geodesic that goes through that event.  From the point of view of the original geodesic, the black hole is a chaotic quantum system in some particular pure state.  To be consistent with this description, the dynamics along the geodesic that goes through the event of horizon crossing, must have kept this subsystem isolated from the rest of the DOF in its Hilbert space, up until that time.  Using the relation between entropy and area, this is equivalent to saying that the system along that trajectory experienced inflation once its apparent horizon size reached $m$.  That is, the model looks, up to this time, like the HST model of dS space.  It has a constant horizon area for a proper time much longer than the size of its apparent horizon.  Our models of cosmology build in homogeneity in the sense that they insist that the sequence of Hamiltonians along each FRW geodesic, is the same.  The consistency conditions require that states also agree on overlaps and the statement that a model of a dilute gas of black holes is a model of inflation is precisely that consistency condition.

To the extent that the dilute black hole gas is uniform out to the largest scales, this means that inflation has to go on for half the conformal time of the entire evolution. Putting in the numbers for the real world, this corresponds to about $60$ e-folds of inflation.  

The HST model describes the traditional ``puzzles" that inflation was invented to solve, in a somewhat novel manner.  The actual quantum model corresponds to a homogeneous cosmological model because of our choice of Hamiltonian, independent of the initial conditions.  The scaling symmetry for large $t$ implies that it's a flat model. It's not clear how to modify it to obtain a model with curved FRW slices.  The horizon problem is clearly solved because the dynamics is built to satisfy causality.  The question of observed homogeneity and isotropy is related to the claim that any alternative configuration of black holes would lead to a universe where the only localized excitations were large black holes and their Hawking radiation.  It's clear that this claim deserves more study.  One should mention that flat anisotropic cosmologies have the topology of a torus.  They would certainly have different consistency conditions between the dynamics along different trajectories, and so constitute a different model of the universe, rather than a different initial condition. The entropy of the Big Bang comes from Hawking decay of tiny black holes, and so there is no ``entropy problem".  

The monopole problem has a novel solution.  There is no era of the universe in which a conventional grand unified field theory is a good description of the physics.  Field theory becomes a good approximation only at temperatures of order $10^{10}$ GeV, way below the unification scale, so monopoles are not formed by the Kibble mechanism.  Using the black hole entropy formula one can estimate the fraction of black holes that are produced with magnetic charge, and it is negligible throughout the history of the universe.

There may be interesting physics associated with the dilute black hole gas era, which precedes the radiation dominated era.  Indeed, the time-scale for primordial perturbations to become non-linear scales according to
\begin{equation} (t/m)^{2/3} \sim m \epsilon , \end{equation} or
\begin{equation} t \sim m^{5/2} \epsilon^{3/2} . \end{equation}  Thus, this happens well before $t \sim m^3$, when the black holes decay.  The universe will thus be filled with primordial structures, whose nature should be calculable in a more or less parameter free way.  These calculations have not yet been done.

If the primordial structures have typical number densities that are everywhere $<< 1/m^3$ then all the black holes will evaporate if they are not magnetically charged.   On the other hand, if the intersections of primordial Zeldovitch pancakes, or the pancakes themselves, become dense enough, then larger black holes, with Hawking temperature lower than the radiation gas, can form.  These black holes will grow by absorption during the radiation and matter dominated eras.   If this leaves over many remnant black holes with lifetime shorter than the age of the universe, the model might be ruled out by our failure to see the Hawking explosions, which terminate the lives of these remnants.   On the other hand, if most of the remnants are cosmologically stable, they could comprise some or all of the dark matter we see in the universe, or could lead to matter domination before it is observed to occur.  Recent surveys\cite{carr} indicate that as much as $10 \%$ of the necessary dark matter could consist of primordial black holes just above the stability bound, and that total rises to $100 \%$ if one discards some of the more uncertain astrophysical bounds.

The reheat temperature is calculated by evaluating the black hole energy density at the time of black hole decay and it is $\sim 10^{10}$ GeV.
 This temperature is equal to the Hawking temperature of a black hole larger than the primordial black hole size by a factor of $g^{1/4} m^{1/2} \sim 5000 $.  It does not seem implausible that primordial black hole mergers of holes of size $m L_P$ could form black holes this large.  Black holes of size larger than this will grow by absorbing radiation during the radiation dominated era and some might grow large enough to have lifetimes longer than the age of the universe (we call this the stability bound). They could then account for some fraction of the observed dark matter.

Clearly one needs to do proper simulations of structure formation in the primeval dilute black hole gas, to elucidate the density and mass distribution of black holes with masses in the vicinity of the stability bound.  It seems implausible that the mechanisms we have outlined could account for black hole masses much larger than the stability bound, and masses much smaller would decay without a trace.   It should then be straightforward to calculate the resulting distribution of black holes at the end of the radiation dominated era, to decide whether the model is ruled out, or the amount of dark matter it can account for.

In addition, we note\cite{tbwfbaryo} that formation of the Hot Big Bang through black hole decay could account for the baryon asymmetry of the universe.  The time variation of the black hole mass provides for CPT violation, so one only needs to understand the origin of CP violation and its role in black hole decays.  The reference cited shows that the numbers required are perfectly plausible.  \section{Conclusions}

The bottom line of this discussion is that the HST model of inflation may provide a complete description of the physics relevant to large scale cosmology, in terms of a small number of parameters already determined by observations of the CMB.  It shares with field theoretic models the need for a slow roll metric, which is used to fit the detailed shape of the CMB power spectrum, but this is really the only free ``parameter" in the model.  Unlike field theory inflation models HST cosmology has at least the potential to explain dark matter and the baryon asymmetry, with no further assumptions.  For baryogenesis, we need a better understanding of the role of CP violation in Hawking decay, and for the hypothetical dark matter model we need N body simulations of the primeval matter dominated era.  The resolution of the monopole problem is completely different from that of conventional inflation models.  Work on all of these topics is in progress.  

It is worth recalling as well that HST inflation is based on a well defined mathematical model of quantum gravity, which is consistent with unitarity and causality (the Hamiltonian is constructed so that the number of variables it couples together increases with the proper time interval along the trajectory whose time evolution it describes).  Conventional inflation models are based on the use of quantum field theory in curved space-time, which has {\it obvious} problems with describing the quantum states near the boundary of a causal diamond.  Field theory models have trouble describing high entropy situations like the production and decay of black holes, while all HST models describe such processes in a completely quantum mechanical framework and do not suffer from ``firewall" paradoxes.  The essence of the firewall paradox\cite{amps} is that the field theoretic claim to model near horizon states in terms of short wavelength modes whose energy is very low in the Hamiltonian used by an asymptotic observer, can only account for that entropy if one claims that an infalling detector is hit by a barrage of short wavelength quanta as it crosses the horizon.  The field theory model also fails to give a microscopic account of the huge increase in black hole entropy that is caused by dropping a light object onto a black hole.  The HST description of such a process\cite{fw3}\cite{gravres14} {\it does} account for the entropy increase and indeed ascribes the long lifetime of the detector inside the horizon to the length of time it takes to excite the extra states, which are made available by the infall.  These degrees of freedom are frozen in the initial state, by the constraints which allow the black hole and the infalling object to behave as independent systems.  The leading large distance  static interaction between localized objects, the Newtonian interaction in the frame where both objects are approximately at rest, comes from virtual excitation of these variables\cite{holonewton}.  During the infall, these variables are excited out of their frozen state, and the time that the infalling object
survives after horizon crossing is the time that it takes to equilibrate these variables (and the object itself) with the majority of the black hole degrees of freedom.

The inflationary era of cosmology is one in which the entropy in a causal diamond is maximal.  The slogan to remember in describing the boundaries of effective quantum field theory is that field theory models fail in situations where the entropy in a space-time region approaches the covariant entropy bound\cite{fsb}.  Low energy density is not sufficient.  Long ago, this was pointed out quite explicitly in \cite{ckn}.  In Minkowski space the states of QFT which do not precipitate the formation of a black hole depend on the region of space-time in which energy density is concentrated in a way that is not captured by short distance cutoffs.  In HST models, the entropy in a region is almost maximal in all situations, but the maximum takes place in the space-time configuration associated with empty space. Localized excitations inside a diamond, are constrained states of the holographic variables that live on the diamond's boundary.  It's only when the number of constrained variables is of order the square root of the total number of variables, that effective field theory gives an approximate description of what's going on. From the HST perspective, field theoretic inflationary models are using effective quantum field theory outside its range of validity.  The use of classical gravitational field theory is however justified, because of Jacobson's demonstration that Einstein's equations are the hydrodynamical equations implied by the area law for entropy. Classical hydrodynamics {\it is} a good description of coarse grained quantities in strongly coupled quantum systems, precisely in states of high entropy.  

The parameters of HST cosmology are the slow roll metric, and the numbers $m$ and $N$.  Apart from the constraint $1 \ll m \ll N$ there does not seem to be anything in the HST formalism that determines these parameters.  Our current data on the inflationary era of the university are not very fine grained.  Any model that gives approximately Gaussian fluctuations that are approximately $SO(1,4)$ invariant, and has a slowly rolling FRW metric gives a coarse grained fit to the data\cite{maldaeftinf}, and the finer details of the scalar two point function are fit by the otherwise unconstrained slow roll metric.  

We need data on tensor fluctuations and non-Gaussianity in order to discriminate between models with wildly different conceptual bases.  Unfortunately, there is a large class of models, probably including HST, in which all of these distinguishing features are predicted to be small.  Finding them in the near term might rule out such models, but if tensor modes and non-Gaussian fluctuations remain elusive, it may be a long time before we can really test models of inflation.  

HST models are distinguished from all other models of early universe cosmology by the fact that they are based on a completely finite mathematical model compatible with many of the requirements of a consistent theory of quantum gravity.  In this essay we've described the models in heuristic terms, in order to make the discussion comprehensible to people unfamiliar with the HST formalism.  The main flaw in HST models is that they do not, in general, satisfy the consistency conditions between the description of physics along trajectories that are in relative motion.  On the other hand, all of the successes of HST models are valid for a huge class of Hamiltonians, so that one may hope that a small subset of them do satisfy those consistency conditions.  If anyone takes up the thankless task of research into HST, this is the problem to solve.

\vskip.3in
\begin{center}
{\bf Acknowledgments }
\end{center}
 The work of T.B. was {\it Not} supported by the Department of Energy.  The contribution of Willy Fischler is based upon work supported by the National Science Foundation under Grant Number PHY-1620610.

%%%%%%%%%%%%%%%%%%%%%%%%%%%%%%%%%%%%%%%%%%%%%%%%%%%%%%%%%%%%%%%%%%%%%%%%%%%%%%%%%%%%%%%%%%%%%%%%%%%%%%%%%%%%%%%%%%%%%%%%%%%%%%%%%%%%%%%%%%%%%%%%%%%%

% \bibliographystyle{utphys}
% \bibliography{fuzzy_refs}

%%%%%%%%%%%%%%%%%%%%%%%%%%%%%%%%%%%%%%%%%%%%%%%%%%%%%%%%%%%%%%%%%%%%%%%%%%%%%%%%%%%%%%%%%%%%%%%%%%%%%%%%%%%%%%%%%%%%%%%%%%%%%%%%%%%%%%%%%%%%%%%%%%%%
%%%%%%%%%%%%%%%%%%%%%%%%%%%%%%%%%%%%%%%%%%%%%%%%%%%%%%%%%%%%%%%%%%%%%%%%%%%%%%%%%%%%%%%%%%%%%%%%%%%%%%%%%%%%%%%%%%%%%%%%%%%%%%%%%%%%%%%%%%%%%%%%%%%%

\end{document}